\documentclass[12pt]{iopart}
\usepackage{iopams}  

\usepackage{graphics}
\usepackage{setstack}
\usepackage{amsfonts}
\usepackage{amssymb}
\usepackage[active]{srcltx}
\usepackage[usenames,dvipsnames]{xcolor}
\usepackage{graphicx,subeqnarray}
\usepackage{dcolumn}
\usepackage{pdfpages}
\usepackage{bm}
\usepackage{soul}
\definecolor{babyblue}{rgb}{0.54, 0.81, 0.94}
\definecolor{rosita}{rgb}{0.97, 0.56, 0.65}
\newcommand{\ket}[1]{\ensuremath{\left|{#1}\right\rangle}}
\newcommand{\bra}[1]{\ensuremath{\left\langle{#1}\right|}}
\newcommand{\braket}[2]{\ensuremath{\langle{#1}|{#2}\rangle}}
\newcommand{\op}[1]{\ensuremath{\hat{\mathnormal{#1}}}}
\newcommand{\beq}{\begin{equation}}
\newcommand{\eeq}{\end{equation}}
\newcommand{\mean}[1]{\ensuremath{\langle{#1}\rangle}}

\begin{document}

\title[]{A new route toward orthogonality}
\author{Andrea Vald\'es-Hern\'andez and Francisco J. Sevilla}

\address{Instituto de F\'isica, Universidad Nacional Aut\'onoma de M\'exico,
Apdo.\ Postal 20-364, 01000, Ciudad de M\'exico, M\'exico}

\eads{\mailto{andreavh@fisica.unam.mx}, \mailto{fjsevilla@fisica.unam.mx}}
\vspace{10pt}
\begin{indented}
\item[]
\end{indented}

\begin{abstract}
We revisit the problem of determining conditions under which a pure state, that evolves under an arbitrary unitary transformation, reaches an 
orthogonal state in a finite amount of the transformation parameter. Simple geometric considerations disclose the existence of a fundamental limit for 
the minimal amount required, providing, in particular, an intuitive hint of the Mandelstam-Tamm bound. The geometric considerations leads us to focus 
on a particular, yet relevant, family of states that evolve towards orthogonality. Several dynamical features are discussed, which include the 
(relative) entropy production during transformation, and special attention is paid to multipartite systems of $N$ bosons that are allowed to tunnel 
between two sites. The effects of the tunneling in the amount of transformation required for the system to attain an orthogonal state are revealed, and 
the relation between the latter, the tunneling intensity and the mode-entanglement is explored.

\end{abstract}

\vspace{2pc}

%

\section{\label{Introduction}Introduction}
\label{intro}

The reliable manipulation of quantum states to reach particular target states lies at the core of quantum information science 
\cite{CanevaPRL2009,DeffnerJPhysB2014}. Particularly, the determination of the \emph{amount of transformation} required to take a specific state 
to an orthogonal ---distinguishable--- one, acquires relevance in quantum information tasks, and has received attention under different contexts, 
mainly in relation with the so-called \emph{quantum speed limit}, in 
which the minimal time required to evolve between two states separated by a given `distance' is considered 
\cite{FlemingNuovoCimento1973,GislasonPRA1985,Mandelstam1991,VaidmanAJP1992,MargolousPhysicaD1998,KosinskiPRA2006,LevitinPRL2009,TaddeiPRL2013,delCampoPRL2013,
DeffnerPRL2013,ZhangSciRep2014,AnderssonJPA2014}. Investigations 
regarding the amount of transformation to take a state into an orthogonal one, are also significant in the broader theoretical and experimental analysis of 
the dynamic evolution of quantum states, in the establishment of fundamental physical limitations (usually expressed as uncertainty relations), as 
well as in other applications of quantum physics \cite{FreyQInfProcc2016}. Moreover, ongoing theoretical analysis on this topic are 
still revealing new aspects on its connection with  several quantum features 
\cite{PiresPhysRevX2016,WeiSciRep2016,JordanPRA2017,DeffnerJPhysA2017,BukovPRX2019,FogartyPRL2020}.

In this paper we revisit the problem of determining the amount of transformation required for an initial pure state to reach an orthogonal one under 
an arbitrary unitary transformation. Our analysis allows to unveil, clarify, and identify different novel aspects related with the dynamics towards 
orthogonality, and adds to preceding works in several ways. In particular, in Section \ref{geo}, we present 
a fresh discussion ---based on associating the overlap between two states with the superposition of two-dimensional rotating vectors
--- 
that brings out the notion of `speed limit' in a most natural geometrical way, and particularly hints out on the Mandelstam-Tamm bound in an intuitive fashion. Further, the amount of evolution (and its lower bound) towards orthogonality is determined for a particular, yet important, family of initial states. This 
amount defines a characteristic scale for which the evolution leads to `clock' models \cite{AnandanPRL1990}, and the dynamics between the `tickings' is analyzed from an entropic perspective. 

In Section \ref{SectIII:application} special attention is paid to multipartite systems of $N$ bosons allowed to tunnel between two sites
(or rather $N$ qubits in an effective Hilbert space of dimension $N+1$). 
The `slowest' and `fastest' states in reaching an orthogonal state are identified. In the non-tunneling regime, a simple relation between the 
amount of transformation needed to attain orthogonality and the states's mode-entanglement is obtained. The effects of the transition between sites 
are also explored, showing that the tunneling favors a `faster' evolution towards orthogonality, and that variations of the tunneling intensity gives 
rise to a rich dynamics of the entanglement between modes (as opposed to the extensively studied case of entanglement between particles). Section \ref{fin} 
presents a summary and some final remarks.

Our results enrich the current comprehension on the dynamics of quantum systems, and in particular of multipartite entangled systems which are of 
extended theoretical and experimental interest \cite{KaufmanScience2016}. 
\section{\label{Shortcut}The geometrical shortcut to orthogonality}\label{geo}

\subsection{The geometrical picture}
We consider a pure-state vector in a Hilbert space $\mathcal{H}$ that evolves under the unitary transformation
\beq \label{Ug}
\op{U}_{g}(\gamma)=e^{-i\op g \gamma},
\eeq
with $\hat g$ an Hermitian operator, and $\gamma$ a real and continuous parameter that defines the $\gamma$-translation symmetry transformation. The 
initial (normalized) state is expanded according to $
 \ket{\psi_0}=\sum^{\mathcal N}_{k=1}\sqrt{r_{k}}e^{i\varphi_{k}}\ket{g_{k}}$,
so the evolved state reads
\begin{equation}\label{Psi-Expansion}
 \ket{\psi_{\gamma}}=\op{U}_{g}(\gamma)\ket{\psi_0}=\sum^{\mathcal N}_{k=1}\sqrt{r_{k}}e^{i(\varphi_{k}-g_k\gamma)}\ket{g_{k}},
\end{equation}
where $\op g\ket{g_k}=g_k\ket{g_k}$, $\varphi_{k}\in [0,2\pi]$, and the coefficients $r_{k}\in[0,1]$ satisfy the normalization condition
\begin{equation}\label{normalization}
 \sum^{\mathcal N}_{k=1}r_{k}=1,
\end{equation}
thus $\{r_k\}$ conforms the probability distribution from which the average $\mean{F(\op{g})}=\sum_{k=1}^{\mathcal{N}}r_{k}F(g_{k})$ 
is computed. This distribution has associated the \emph{information} (Shannon) \emph{entropy}  
$S_{\rm Sh}[\{r_{k}\}]=- \sum^{\mathcal N}_{k=1}r_{k}\ln r_{k}$, which measures the lack of information about the state 
$\ket{\psi_{\gamma}}$ relative to the basis $\{\ket{g_{k}}\}$. Its minimum value ($S_{\rm Sh}=0$) 
is attained when there is maximal information, corresponding to $r_{k}=\delta_{k,k_0}$, whereas its 
maximum value ($S_{\rm Sh}=1$) occurs when the available information is minimal, so the 
distribution is $r_k=1/\mathcal{N}$, in line with the \emph{equal a priori probability} principle.

One quantity of interest that characterizes the evolution of pure states is 
the overlap 
\begin{equation}\label{Overlap}
 \braket{\psi_{\gamma}}{\psi_{0}}=\sum^{\mathcal N}_{k=1}r_{k}e^{ig_{k}\gamma}=\mean{e^{i\op{g}\gamma}},
\end{equation}
which is a $\gamma$-dependent complex number that measures how similar is the transformed state $\ket{\psi_{\gamma}}$ 
with respect to the original one, $\ket{\psi_{0}}$. \Eref{Overlap} vanishes provided $\ket{\psi_{\gamma}}$ is orthogonal to 
(distinguishable from) $\ket{\psi_{0}}$. A \emph{relative entropy} between such 
states can be defined in terms of (\ref{Overlap}) as  $S_{\rm 
rel}\big[\{r_{k}\};\gamma\big]=1-\left\vert\braket{\psi_{\gamma}}{\psi_{0}}\right\vert^{2
}$, with $\left\vert\braket{\psi_{\gamma}}{\psi_{0}}\right\vert^{2}$ known as the
\emph{non-decay} or 
\emph{survival probability} whenever $\gamma$ and $\op{g}$ are identified, respectively, with time and the 
Hamiltonian operator \cite{GislasonPRA1985} (we take $\hbar=1$). The relative entropy 
acquires its maximum ($S_{{\rm rel}}=1$) and minimum ($S_{\rm rel}=0$) value when the 
transformed state is orthogonal and equal (up to a phase factor) to the original one, respectively.

Each term of the expansion in (\ref{Overlap}) can be mapped into the two-dimensional vector $\boldsymbol{r}_k(\gamma)=r_k(\cos \theta_k,\sin 
\theta_k)$ lying inside the unitary circle ($r_k\leq1$) with $\theta_k=g_k\gamma$. Consequently $\braket{\psi_\gamma}{\psi_0}$ is mapped into the 
vector $\boldsymbol{R}(\gamma)= \sum^{\mathcal N}_{k=1}\boldsymbol{r}_k(\gamma)$, and thus the overlap vanishes if 
and only if 
$\boldsymbol{R}=\boldsymbol{0}$. This observation offers a simple geometrical prescription to 
guarantee that $\ket{\psi_{0}}$ evolves towards an orthogonal state, namely: sum up the vectors 
$\boldsymbol{r}_k(\gamma)$, each performing uniform circular motion with frequency $g_k=d\theta_k/d\gamma$ 
along orbits of radius $r_k$, and from this sum, determine the conditions under which the vector's tips are distributed such that the centroid of the 
polygon they form coincides with the origin (this is the case for example, when the tips form a regular 
polygon centered at the origin). From this geometrical picture it follows 
immediately that: 
\begin{itemize}
\item[(a)] The number of vectors $\boldsymbol{r}_{k}$ equals $\mathcal{N}$, the number of states that contribute in the expansion 
(\ref{Psi-Expansion}). 
\item[(b)] Each of the $\mathcal{N}$ phase oscillators is accommodated in one of the $\mathcal{R}$ different radii ($1\leq \mathcal{R}\leq 
\mathcal{N}$), with $\mathcal{R}$ the number of different coefficients $r_k$.
\item[(c)] The number of frequencies equals the number of different eigenvalues $g_k$ appearing in (\ref{Psi-Expansion}). \label{iii}
\end{itemize}

\begin{figure}
\centering
 \includegraphics[width=0.40\textwidth]{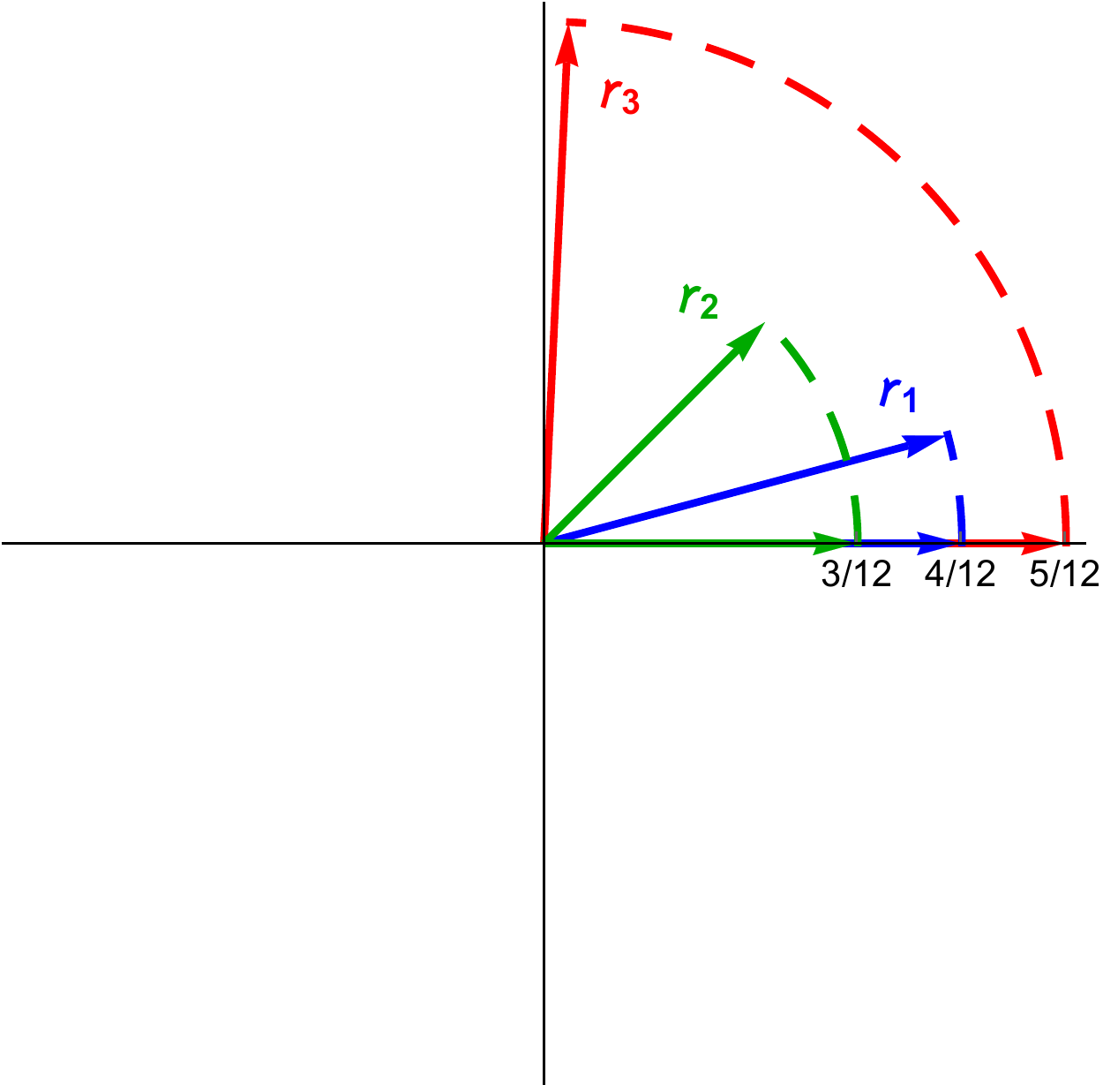}\quad\qquad\includegraphics[width=0.40\textwidth]{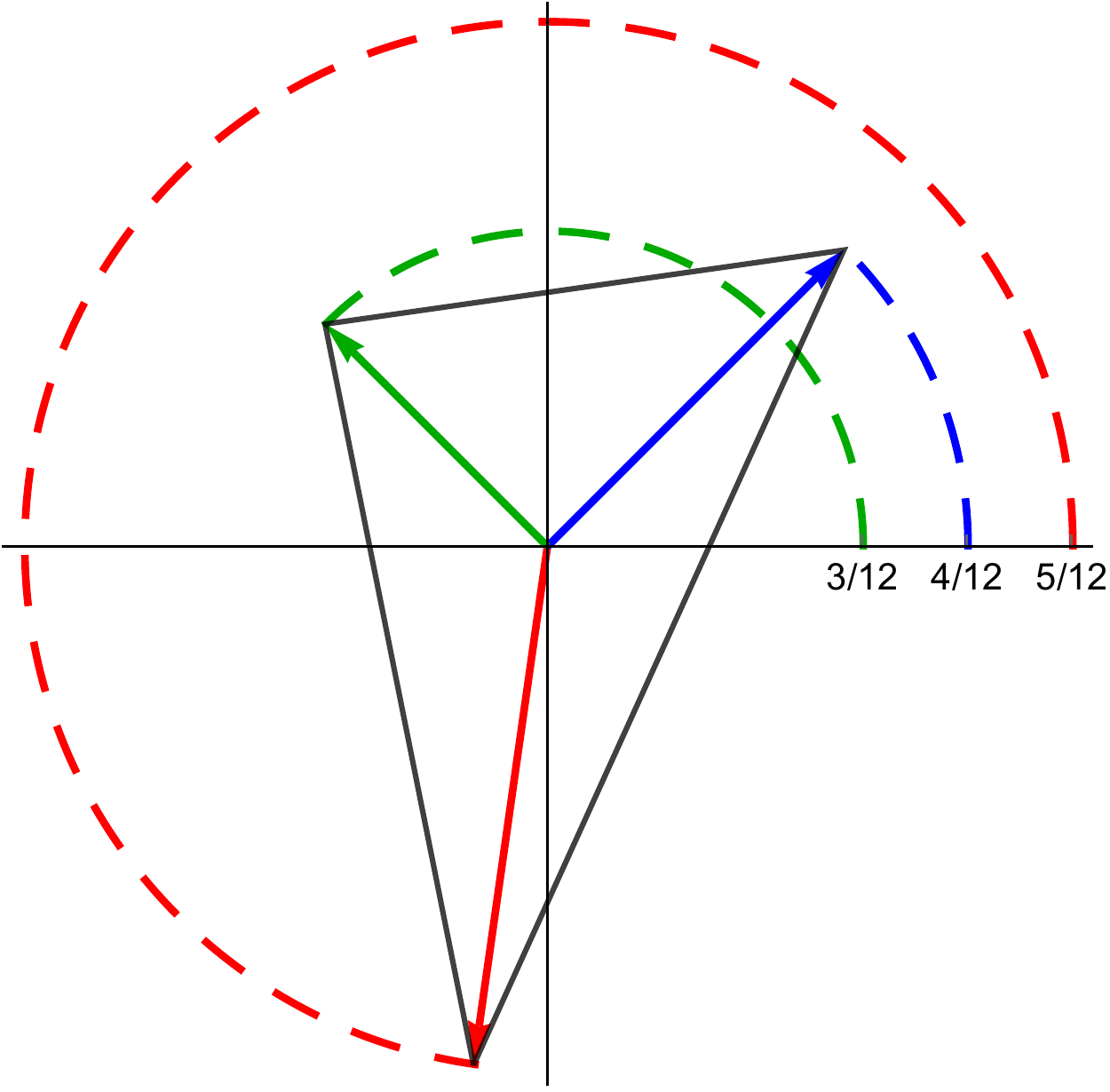}
 \caption{\label{FigRotation}Left panel: Initially ($\gamma=0$) all $\boldsymbol{r}_k$ ($k=1,2,3$) lie along the horizontal axis, and for $\gamma>0$ they perform a counter-clockwise rotation with $\theta_k=g_k\gamma$. For $0<\gamma<\widetilde{\gamma}$ clearly $\boldsymbol{R}(\gamma)= \sum^{3}_{k=1}\boldsymbol{r}_k(\gamma)$ is not a vanishing vector. Right-panel: At $\gamma=\widetilde{\gamma}$ the tips of the vectors form a polygon with its centroid at the origin, hence the vector $\boldsymbol{R}(\widetilde\gamma)$ vanishes, indicating that the system has reached a state orthogonal to $\ket{\psi_0}$.}
\end{figure}

The initial state is depicted such that all the vectors $\boldsymbol{r}_{k}(0)$ lie on the $x$-axis at a distance 
$r_k$ from the origin. When the evolution is switched on, each vector starts its 
counter-clockwise rotation, and for a given $\gamma>0$ they are all distributed inside the unitary circle in positions that 
depend on their frequency. 
Consequently, it is intuitively clear that for 
all these vectors to sum up to a null vector (see figure \ref{FigRotation} for an example with $\mathcal N=3$): 
\begin{itemize}
\item[d)] At least two vectors with different frequencies, i.e., two \emph{different} eigenvalues $g_k$,  must contribute in (\ref{Psi-Expansion}).

\item[e)] The value of $\gamma$ at which this occurs, denoted as $\widetilde{\gamma}$, must be larger than or equal to a certain \emph{minimum} value $\widetilde{\gamma}_{\min}$, that is, 
$\widetilde{\gamma} \geq \widetilde{\gamma}_{\min} 
>0$.    
\end{itemize}

From d) it follows that a nonzero value of the ($\gamma$-invariant) dispersion $\sigma^2_{\op 
g}=\langle \op g^2\rangle-\langle \op g\rangle^2$ is necessary for orthogonality to be attained. 
Since this quantity is equivalent to the variance of the eigenvalues $\{g_k\}$ computed with 
the distribution $\{r_{k}\}$, the latter plays a fundamental role in attaining the orthogonal 
state. In particular, a distribution $\{r_{k}\}$ with nonzero $S_{\rm Sh}$ 
is required for orthogonality to be reached. On the other hand, e) implies that if  
$\braket{\psi_\gamma}{\psi_0}$ vanishes, then a sort of \emph{speed limit} ---or $\gamma$-$\op{g}$ 
\emph{uncertainty relation}--- exists, determined only by $\widetilde{\gamma}_{\min}$. When the unitary transformation (\ref{Ug}) refers to the temporal evolution, that is, when $\op 
g$ and $\gamma$ correspond, respectively, to a time-independent Hamiltonian and to time, these intuitive conditions become 
formally stated and quantified in the form of the Mandelstam-Tamm bound for 
the so-called \textit{quantum speed limit}, or the time-energy uncertainty relations \cite{GislasonPRA1985,Mandelstam1991}. 

Notice that according to the present geometrical argument, the existence of a speed limit originates in the fact that the dynamics of each term in the overlap (\ref{Overlap}) is performed (uniformly) on a circle, and requires no explicitly quantum hypothesis. Interestingly, speed limits based on the dynamics on the circle \cite{WinfreeBook} can also found, for example, in the emergence of collective synchronization of interacting phase oscillators \cite{TimmePRL2004}. These observations are pertinent in relation with recent works that point towards an ubiquitous nature of the limitations in the speed of evolution, by analyzing the existence of speed limits in the classical realm 
\cite{ShanahanPRL2018,OkuyamaPRL2018,ShiraishiPRL2018}.

\subsection{Lower bound for the `speed' limit}\label{speedlimit}
The problem of determining the conditions on the distribution $\{r_k\}$ and the set $\{g_k\}$ 
for $\ket{\psi_0}$ to reach an orthogonal state at some finite $\widetilde{\gamma}$ is an interesting one \cite{AnderssonJPA2014,HegerfeldtPRL2013}, yet it represents a difficult task in the 
most general case. However, we can make progress either by assuming a particular distribution 
$\{r_{k}\}$ and determining the requirements on $\{g_k\}$, or the other way around. We 
shall adopt the first strategy and consider a distribution $\{r_{k}\}$ that fullfills the \emph{equal a priori probability} principle. Thus, in what 
follows we will consider the family $\mathcal{I}$ of initial states that are equally-weighted superpositions of eigenvectors of $\op{g}$, 
i.e., the collection of states of the form
\begin{equation}\label{EqWeighted-InitialSate}
 \ket{\psi_{0}}=\frac{1}{\sqrt{\mathcal{N}}}\sum_{k=1}^{\mathcal{N}}e^{i\varphi_{k}}\ket{g_{k}}
\end{equation}
where we have chosen $r_k=1/\mathcal{N}$, with finite $\mathcal{N}\geq 2$ (the case $\mathcal{N}$=1 clearly never reaches 
an orthogonal state). This is a natural choice for the distribution $\{r_{k}\}$, assuming we have the minimal information, hence maximal Shannon 
entropy.
We further assume that all the $g_k$ appearing in the expansion (\ref{Psi-Expansion}) are different, and that (without loss of generality) the index 
$k$ is such that $g_k<g_{k+1}$. According to (b) above, this means that the tips of the $\mathcal{N}$ vectors lie on the circle of radius 
$1/\mathcal{N}$, they 
rotate with different angular frequencies $g_{1}<\cdots<g_{k}<\cdots<g_{\mathcal{N}}$, and after a transformation by $\gamma$ units they point at the 
directions $\theta_1<\cdots<\theta_k<\cdots<\theta_\mathcal{N}$, respectively. 

\begin{figure}[h]
\qquad\qquad\includegraphics[width=0.34\textwidth]{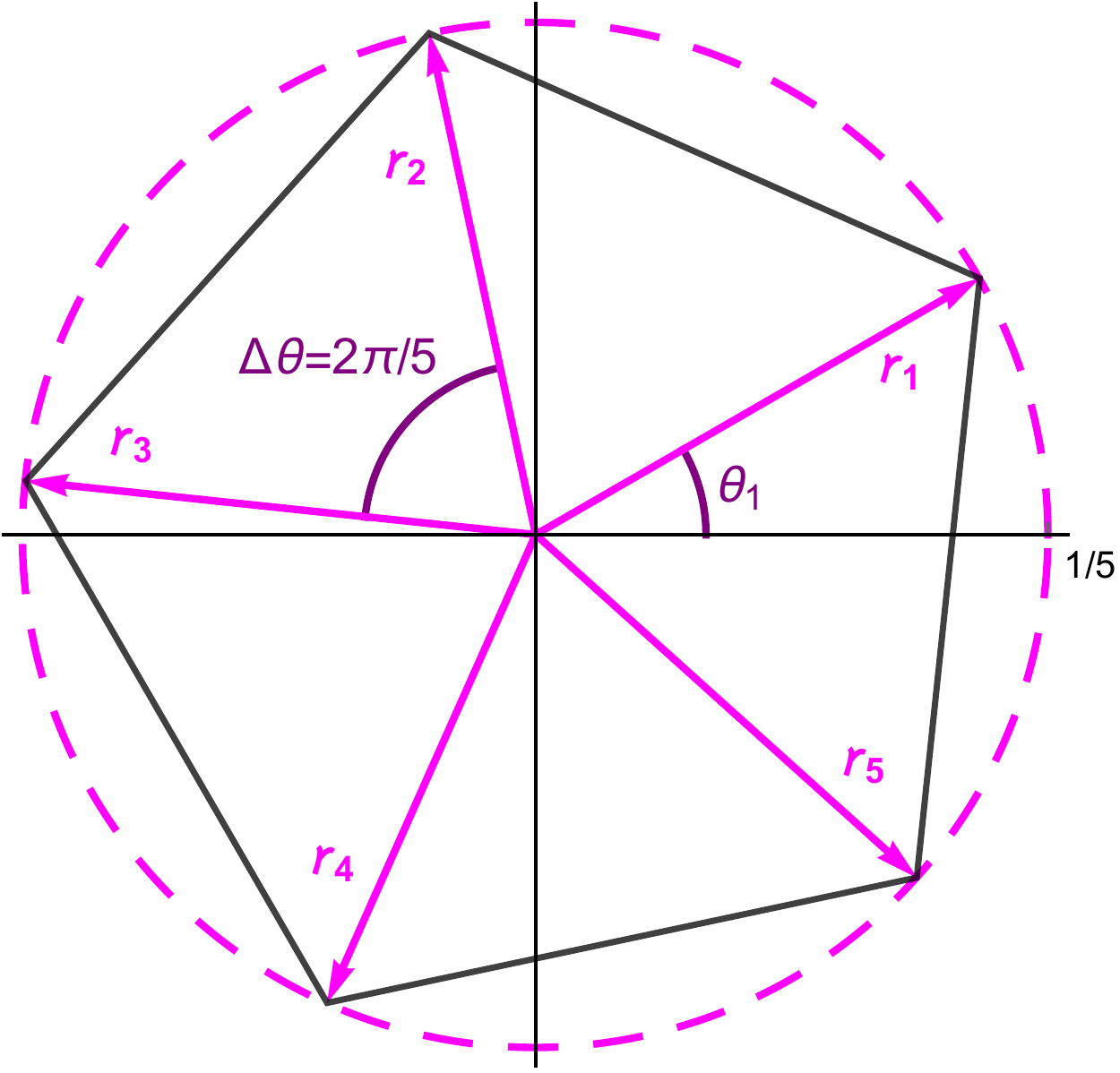}
 \caption{\label{pentagon}An equally weighted superposition of 5 states reaches an orthogonal state as soon as the rotating vectors $\boldsymbol r_{k}$ ($k=1,\dots,5$) form a regular pentagon. The angle $\theta_1$ is given by $\theta_1=g_1\widetilde\gamma=2\pi g_1/5\Delta g$, as follows from equation (\ref{gammapoli}).}
\end{figure}

From this geometrical description, it is clear that an orthogonal state is reached \emph{for the first time} at 
$\widetilde\gamma$ ---that is, $\braket{\psi_{\widetilde{\gamma}}}{\psi_{0}}=0$ or equivalently, 
$\boldsymbol{R}\,(\widetilde\gamma)=\boldsymbol{0}$--- whenever the tip vectors form (for the first time) the $\mathcal{N}$ vertices 
of a regular polygon inscribed in the circle of radius $1/\mathcal N$, as illustrated in figure \ref{pentagon} for $\mathcal N=5$. In this case it holds that $\Delta\theta=\theta_{k+1}-\theta_{k}=2\pi/\mathcal{N}$, meaning that
\footnote{The condition of the regular polygon is sufficient, although 
not necessary to attain orthogonality; however, for $\mathcal N=2$ it is a necessary and sufficient condition. Indeed, for $\mathcal N=2$ there are 
only two vectors that perform phase rotation, and for these to sum up a null vector both radii must necessarily 
be equal (i.e., $r_1=r_2=1/2$), and one vector opposite to the other forming a diameter of the circle (the `polygon' 
in this case). Consequently, 
an equally weighted \emph{qubit state} always reaches an orthogonal state at 
$\widetilde{\gamma}=\pi/(g_2-g_1)$, as follows from Eq. (\ref{gammapoli}).}
\numparts
\beq \label{gammapoli}
\widetilde{\gamma}=\frac{2\pi}{\Delta g}\frac{1}{\mathcal{N}},
\eeq
where
\beq\label{Deltag}
\Delta g=g_{k+1}-g_k 
\eeq
\endnumparts
is independent of $k$, which occurs only if the elements of the set $\{g_{k}\}$ are of 
the form 
\beq \label{spec}
g_k=(k-1)\Delta g+g_1.
\eeq
With this, all the $g$-distributed  moments, $\mean{\op{g}^{n}}= \sum^{\mathcal N}_{k=1}r_{k}g_{k}^{n}=\mathcal{N}^{-1}\sum^{\mathcal 
N}_{k=1}g_{k}^{n}$, (which are not all independent among themselves) are explicitly 
determined in terms of $g_1$, $\mathcal{N}$, and $\Delta g$ only. Thus, different, yet equivalent, expressions for $\widetilde{\gamma}$ are 
possible when this is written in 
terms of any of the moments of $\op{g}$. Of particular interest are those expressions for $\widetilde{\gamma}$ that involve the first 
two moments \cite{LevitinPRL2009,ZielinskiPRA2006}
\numparts
\begin{eqnarray}
   \mean{\op{g}}&=\frac{1}{\mathcal{N}}\sum_{k=1}^\mathcal{N} g_k=g_1+\frac{\Delta 
g}{2}(\mathcal{N}-1),\label{Epoli}\\
    \mean{\op{g}^{2}}&=\frac{1}{\mathcal{N}}\sum_{k=1}^\mathcal{N} g_k^{2}=g_{1}^{2}+g_{1}\Delta g\, 
(\mathcal{N}-1)+\frac{1}{6}(\Delta g)^{2}\, (\mathcal{N}-1)(2\mathcal{N}-1),\label{SigmaPoli} 
\end{eqnarray}
\endnumparts
since these play the most determinative role in the transformation (\ref{Ug}) \cite{LevitinPRL2009,AnandanPRL1990}.

In terms of $\mean{\op{g}}-g_1$ ---the mean value of $\op g$ as measured from the lowest eigenvalue appearing in equation (\ref{EqWeighted-InitialSate})---, 
equation (\ref{gammapoli}) rewrites as 
\beq \label{gammapoli2}
\widetilde{\gamma}=\frac{\pi}{2(\mean{\op{g}}-g_1)}\frac{2(\mathcal{N}-1)}{\mathcal{N}}\geq \frac{\pi}{2(\mean{\op{g}}-g_1)},
\eeq
whereas in terms of the standard deviation 
$\sigma_{\op{g}}=\sqrt{\mean{(\op{g}-\mean{\op{g}})^{2}}}=\sqrt{\mean{\op{g}^{2}}-\mean{\op{g}}^{2}}$, $\widetilde{\gamma}$ can be written as
\begin{equation}\label{gsigma} 
\widetilde{\gamma}=\frac{\pi}{2\sigma_{\op{g}}}\sqrt{\frac{4\,(\mathcal{N}^{2}-1)}{
3\,\mathcal{N}^{2}}}\geq \frac{\pi}{2\sigma_{\op{g}}}.
\end{equation}
The inequalities in equations (\ref{gammapoli2}) and (\ref{gsigma}) are, respectively, the extension to the $\gamma$-translation transformation of the 
Margolus-Levitin \cite{MargolousPhysicaD1998} and the Mandelstam-Tamm \cite{Mandelstam1991} bounds that determine 
the quantum speed limit. In particular, equation. (\ref{gsigma}) leads to the inequality $
\sigma_{\op{g}}\widetilde{\gamma}\ge \pi/2$,
which plays the role of a $\gamma$-$\op{g}$ `uncertainty' relation \cite{GislasonPRA1985}. Moreover, the quantity 
$s=2\sigma_{\op{g}}\widetilde{\gamma}$ is independent of the particular generator $\op{g}$ that defines the unitary transformation (\ref{Ug}), and thus 
corresponds to the analogous of a geometric phase as discussed in reference \cite{AnandanPRL1990}. 

The bounds indicated in equations (\ref{gammapoli2}) and (\ref{gsigma}) have been synthesized in the 
following expression for the lower bound of $\widetilde{\gamma}$ \cite{Giovannetti2003EPL,{Giovannetti2003PRA}}
\beq\label{min}
\widetilde{\gamma}_{\min}=\max\Big\{\frac{\pi}{2(\mean{\op{g}}-g_1)},\frac{\pi}{2\sigma_{\op{g}}}\Big\}.
\eeq
The quantities inside the brackets are not necessarily independent of each other, and as shown by Levitin and Toffoli \cite{LevitinPRL2009}, the bound 
(\ref{min}) can be attained only when $(\mean{\op{g}}-g_1)=\sigma_{\op{g}}$, and is asymptotically attainable otherwise.

Now, the equalities in equations (\ref{gammapoli2}) and (\ref{gsigma}) lead to
\beq\label{quot}
\frac{\sigma_{\op{g}}}{(\mean{\op{g}}-g_{1})}= \sqrt{\frac{1}{3}\frac{(\mathcal{N}+1)}{(\mathcal{N}-1)}}\leq 1,
\eeq 
thus disclosing the relation between $\mean{\op{g}}-g_1$ and $\sigma_{\op{g}}$ (which is of course particular for the specific distribution 
$\{r_{k}\}$ under consideration), and accordingly for the family of initial states $\mathcal{I}$ we have
\beq\label{min2}
\widetilde{\gamma}_{\min}=\frac{\pi}{2\sigma_{\op{g}}}.
\eeq

The quotient in equation (\ref{quot}) approaches  $1/\sqrt{3}\approx0.57735$ as $\mathcal{N}\rightarrow\infty$ meaning that the fluctuations 
are not negligible even in the regime of large systems. On the other hand, the quotient in (\ref{quot}) saturates the inequality for $\mathcal{N}=2$, 
and it is only for this value of $\mathcal N$ that the Margolus-Levitin and the Mandelstam-Tamm bounds coincide and $\widetilde{\gamma}$ attains its minimum value, as follows from equation (\ref{gsigma}) (thus 
recovering the main 
statement established in theorem 1 of reference \cite{LevitinPRL2009}). Therefore, for fixed $\mean{\op{g}}-g_1$ (or $\sigma_{\op{g}}$), a 
superposition involving two states only (an effective qubit), reaches an orthogonal state faster than any 
other superposition. In contrast, for the same $\mean{\op{g}}-g_1$ we have that $\sigma_{\op{g}}<\mean{\op{g}}-g_{1}$ for $\mathcal{N}>2$, and thus
$\widetilde{\gamma}$ asymptotically approaches the value $(2/\sqrt{3})\widetilde{\gamma}_{\rm min}$ with  
$\mathcal{N}\rightarrow \infty$.

Throughout the paper, when we say that a given state evolves `faster' than other one, we are tacitly considering the \emph{absolute} amount of evolution $\gamma$ (a universal parameter that characterizes the pace of change of \emph{all} systems' states on equal footing), so that a state $\ket{\psi_a}$ is said to reach orthogonality faster than $\ket{\psi_b}$ whenever $\widetilde{\gamma}_{a}<\widetilde{\gamma}_{b}$. However, it should be bear in mind that other \emph{state-dependent} scales can be considered, leading to different and relative concepts of speed of evolution that are also useful in characterizing the dynamics of the system. One example is the ratio $\mathsf{g}=\gamma/\widetilde{\gamma}_{\min}$,
which quantifies the amount of evolution relative to the minimal amount imposed by the system's own dynamical limitations. In this case, $\ket{\psi_a}$ is said to be faster than $\ket{\psi_b}$ if $\widetilde{\mathsf{g}}_{\,a}<\widetilde{\mathsf{g}}_{\,b}$, with 
\begin{equation}\label{ratio}
\widetilde{\mathsf{g}}=\frac{\widetilde \gamma}{\widetilde{\gamma}_{\min}}\geq1.
\end{equation}
Another notion of speed of evolution ensues by focusing on the quantity $\widetilde{\gamma}_{\min}$,  according to which the faster state is that with the minimum quantum speed limit. Notice, however, that this figure of merit does not necessarily compare the amount of transformation required to actually reach orthogonality, because (as stated above) the bound $\widetilde{\gamma}_{\min}$ is in general not attainable.   

Since $\widetilde{\gamma}_{\min}$ varies in general from state to state, it may occur that $\widetilde{\gamma}_{a}<\widetilde{\gamma}_{b}$ but $\widetilde{\mathsf{g}}_{\,b}<\widetilde{\mathsf{g}}_{\,a}$, so clearly the notion of `faster' depends on the quantity employed to measure the speed of evolution. In particular, for the family of states considered above, equations (\ref{gsigma}) and (\ref{min2}) lead to    
\begin{equation}\label{ratio2}
\widetilde{\mathsf{g}}=\frac{2}{\sqrt 3}\sqrt{\frac{\mathcal{N}^{2}-1}{
\mathcal{N}^{2}}},
\end{equation}
so $\widetilde{\mathsf{g}}$ depends only on $\mathcal N$, whereas $\widetilde{\gamma}$ depends on $\mathcal N$ and also on the separation $\Delta g$, as seen from equation (\ref{gammapoli}). Therefore, for example, two states $\ket{\psi_a}$ and $\ket{\psi_b}$ having the same number of terms in their expansion (\ref{EqWeighted-InitialSate}) are equally fast in terms of $\mathsf{g}$, yet in terms of the absolute parameter $\gamma$ the state $\ket{\psi_a}$ is faster provided $\Delta g_b<\Delta g_a$, that is, provided the separation between its states is larger.

\subsection{Evolution between orthogonal states}
Besides determining the value for which the transformed state is orthogonal to the 
initial one, $\widetilde{\gamma}$ also defines a collection of $\mathcal 
N$ mutually orthonormal states. First notice that for integers
$n,m=0,1,\dots,\mathcal{N}-1$, the evolution operator (\ref{Ug}) implies
that
\beq \label{nm}
\langle\psi_{n\widetilde{\gamma}}|\psi_{m\widetilde{\gamma}}\rangle=
\langle\psi_{(n-m)\widetilde{\gamma}}|\psi_{0}\rangle= \sum^{\mathcal N}_{k=1}r_{k}e^{ig_{k}(n-m)\widetilde \gamma}.
\eeq 
In the present case we have $r_k=1/\mathcal N$, $g_k$ is given by equation (\ref{spec}), and 
$\widetilde\gamma$ is given by equation (\ref{gammapoli}). Gathering these results equation (\ref{nm}) becomes
\beq \label{nm2}
\langle\psi_{n\widetilde{\gamma}}|\psi_{m\widetilde{\gamma}}\rangle=
e^{ig_{1}(n-m)\widetilde \gamma}\,\frac{1}{\mathcal N}\,\sum^{\mathcal{N}-1}_{l=0}e^{2\pi 
i\frac{1}{\mathcal{N}}(n-m)l}=e^{ig_{1}(n-m)\widetilde \gamma}\,\delta_{nm}=\delta_{nm},
\eeq 
meaning that $\mathcal{O}=\{\ket{\psi_{0}},\ket{\psi_{\widetilde{\gamma}}},\dots,\ket{\psi_{(\mathcal{N}-1)\widetilde{\gamma}
}}\}$ is indeed a set of $\mathcal N$ mutually orthonormal states, obtained by the successive 
application: once, twice, \ldots, $\mathcal {N}-1$ times, of the operator $\hat U_g(\widetilde\gamma)$ to the initial 
state $\ket{\psi_0}$. From equation (\ref{nm2}) we have that the 
relative entropy $S_{\rm rel}$ between any pair of these states acquires its maximum value 1.
\begin{figure}
\centering
 \includegraphics[width=0.75\textwidth]{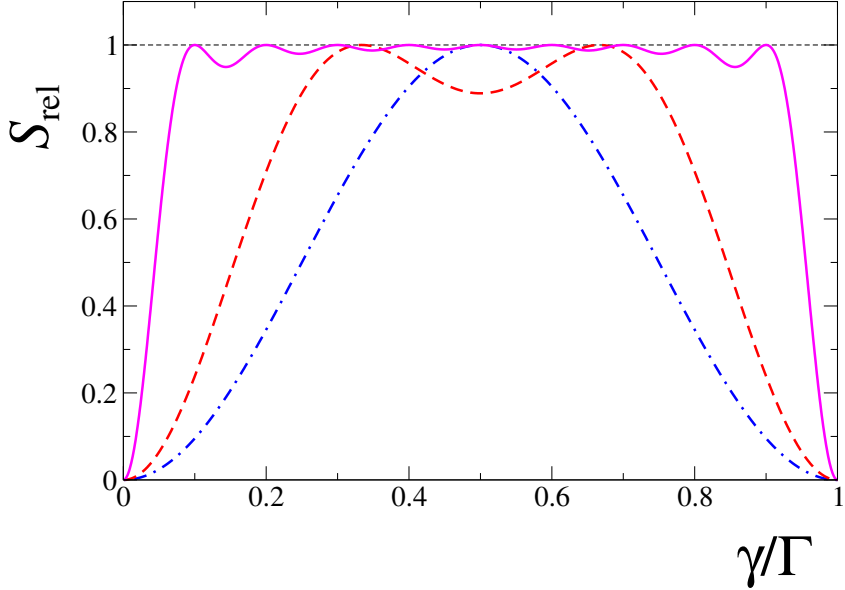}
 \caption{\label{Fig:RelEnt}The relative entropy $S_{\rm rel}=1-|\langle\psi_\gamma|\psi_0\rangle|^2$ as a function of $\gamma/\Gamma$, with $\Gamma=\mathcal{N}\widetilde{\gamma}$, for $\mathcal{N}=2$ (blue, dash-dotted line), 3 (red, dashed line) and 10 (pink, solid line). $S_{\rm rel}$ attains its maximum value $1$ each time an orthogonal state is reached, and becomes $0$ only when the initial state is recovered, at $\gamma=\mathcal{N}\widetilde{\gamma}$. For sufficiently large $\mathcal N$, the state $\ket{\psi_\gamma}$ is highly distinguishable from $\ket{\psi_0}$ in the entire interval $\widetilde{\gamma}\leq\gamma\leq(\mathcal N-1)\widetilde{\gamma}$. }
\end{figure}

Figure \ref{Fig:RelEnt} shows the relative entropy between $\ket{\psi_0}$ and $\ket{\psi_\gamma}$ for three 
different initial states, corresponding to $\mathcal{N}=2$, 3 and 10 (for comparison purposes $S_{\rm rel}$ is plotted as a function of 
$\gamma/\Gamma$ with $\Gamma=\mathcal{N}\widetilde{\gamma}$). As seen in the figure, the relative entropy is produced in the passing from the initial state ($\gamma=0$) 
to the first orthogonal state ($\gamma=\widetilde\gamma$), when $S_{\rm rel}$ attains its maximum value for the first time. Then $S_{\rm rel}$ exhibits oscillations reaching its maximum value periodically, as $\gamma$ take the values $2\widetilde\gamma,\dots,(\mathcal N-1)\widetilde\gamma$, that is, each time the system visits an element of $\mathcal O$. From that point on, the relative entropy is degraded as the state goes back to the initial state ($\gamma=\mathcal N\widetilde\gamma$), as expected by the recurrent nature of the dynamics.
Remarkably, for $\mathcal{N}>2$, there is little variation of the relative entropy production when the system evolves between successive orthogonal states ($S_{\rm rel}$ remains close to 1), meaning that $\ket{\psi_{\gamma}}$ remains `far'
from $\ket{\psi_{0}}$ almost all time, a feature that gets conspicuous as $\mathcal{N}$ gets larger.

The basis $\mathcal{O}$ is encountered, for instance, in the theory of a one-dimensional \emph{locally periodic media} \cite{GriffithsAJP2000}, 
in which case the $n$th element of $\mathcal{O}$ corresponds to the state of a particle in the $n$th cell of the system, the unitary transformation 
corresponds to the spatial translation, and $\widetilde{\gamma}$ to the length of the unit cell.
When $\op{U}_{g}(\gamma)$ corresponds to the time-translation transformation, the states $\{\ket{\psi_{m\widetilde{\gamma}}}\}$ coincide with the 
pointer positions of the simple model of the quantum clock proposed by Anandan and 
Aharonov in reference \cite{AnandanPRL1990}. In our geometrical picture, that associates the condition 
$\langle\psi_{\widetilde{\gamma}}|\psi_0\rangle=0$ with a regular polygon with $\mathcal{N}$ 
(labeled) vertices, each application of $\hat U_g(\widetilde\gamma)$ amounts to perform a 
counterclockwise rotation that takes one vertex into the next one (in the time-translation case, takes the hands of the clock into each `second' 
positions). Thus, at each rotation by the angle $\widetilde{\gamma}\Delta g $, 
an element of the basis $\{\ket{\psi_{n\widetilde{\gamma}}}\}$ is reached. Clearly the $\mathcal 
N$th rotation brings back the vertices to their original position, meaning that the evolution is 
cyclic, with period 
\begin{equation}\label{PeriodGamma}
 \Gamma=\mathcal{N}\widetilde{\gamma}=2\pi/\Delta g,
\end{equation}
a quantity that has been identified as a characteristic time scale in the context of time evolution 
\cite{MajteyEntropy2019}. 

With the aid of Eq. (\ref{spec}), a direct calculation shows that indeed, up to a global ($g_1$-dependent) phase factor, 
the following condition holds 
\begin{eqnarray}
 \label{cond}
\ket{\psi_\gamma}&=\op{U}_g(\Gamma)\ket{\psi_\gamma}\nonumber\\
&=\left[\op{U}_{g}(\widetilde{\gamma})\right]^{\mathcal{N}}\ket{\psi_\gamma},
\end{eqnarray}
or equivalently,
\beq\label{cond2}
\op{U}_g(\gamma)\ket{\psi_0}=\op{U}_g(\gamma+\Gamma)\ket{\psi_0}.
\eeq
Consequently, the state $\ket{\psi_\gamma}$ is invariant under the single-period  transformation 
$\op{U}(\Gamma)$. When $\op{U}$ corresponds to the time-evolution operator, $\op{U}(\Gamma)$ is 
known as the \emph{Floquet operator} \cite{BerdanierPNAS2018}, and equation (\ref{cond2}) is consistent 
with the \emph{quantum recurrence theorem} \cite{BocchieriPR1957}.

Before ending this section it is important to make some remarks on the applicability of the previous analysis. Two basic assumptions regarding the initial state $
 \ket{\psi_0}=\sum^{\mathcal N}_{k=1}\sqrt{r_{k}}e^{i\varphi_{k}}\ket{g_{k}}$ were made; the first one fixed $r_k=1/\mathcal N$, and the second one imposed the (equally spacing) condition (\ref{Deltag}) on the eigenvalues $g_k$, namely that $\Delta g=g_{k+1}-g_k$ is constant for all $k\leq\mathcal{N}-1$. It should be stressed, however, that this is \emph{not} a condition imposed on the spectrum of $\hat g$, but rather a restriction on the particular $\hat g$-eigenstates that appear in the expansion of $
 \ket{\psi_0}$. That is, the spectrum of $\hat g$ \emph{needs no} to be an equally-spaced one (as, i.e., in the harmonic oscillator case), yet all the (equally probable) \emph{contributing} eigenstates must comply with the restriction $\Delta g=\rm{constant}$.\footnote{For example, for an energy spectrum of the form $E_k=E_1k^2$, with $k$ a positive integer (particle inside an impenetrable box), or $E_k=E_1k^{-1}$ (hydrogenoid atom), it is not difficult to chose a set of eigenstates that comply with condition (\ref{Deltag}).} In its turn, it is clear that $\{r_k\}$ is a uniform distribution only in the restricted subspace of eigenstates that satisfy condition (\ref{Deltag}), though in the complete space of all $\hat g$-eigenstates it is highly non-uniform. 
 

\section{\label{SectIII:application}An application: $N$-two-level systems}
Two-level, or qubit systems, have proven to be useful in shading light on some aspects of the quantum speed limit \cite{TaddeiPRL2013,HegerfeldtPRL2013,DehdashtiPRA2015,ZhangPLA2018,BasonNPhys2012}. Here we will consider multi-qubit systems, and apply the above results to get further insight into the dynamics of systems composed of $N$ bosons in two sites (which are equivalent to $N$ qubits in a Hilbert space with an effective dimension equal to $N+1$).

\subsection{The generator of unitary transformations}
When describing a system of $N$ bosons in two modes, it is usual to resort to the standard annihilation ($\op{a}_{i}$) and 
creation ($\op{a}_{i}^{\dagger}$) operators of a particle occupying the level $i=0,1$. These operators define the number operator 
$\op{n}_{i}=\op{a}_{i}^{\dagger}\op{a}_{i}$ describing the number of particles in the level $i$, and satisfy the usual commutation relations
\beq \label{comm}
[\op{a}_{i},\op{a}^{\dagger}_{j}]=\delta_{ij}, \quad
[\op{a}_{i},\op{a}_{j}]=0.
\eeq
In terms of the $\op{a}$'s
operators, 
the generator $\op{g}$ of unitary transformations can be written as
\begin{equation} 
\op{g}=\sum_{i,j=0,1}G_{ij}\op{a}_{i}^{\dagger}\op{a}_{j}+\sum_{i,j,k,l=0,1}G_{ijkl}\op{a}^{\dagger}
_{i}\op{a}^{\dagger}_{j}\op{a}_{k}\op{a}_{l}+\ldots,
\end{equation}
where the $G$'s are real parameters. Here we will consider the simpler structure
\begin{equation}\label{1BodyG}
\op{g}=G_{0}\op{n}_{0}+G_{1}\op{n}_{1}+G_{01}\bigl(\op{a}_{0}^{\dagger}\op{a}_{1}
+\op { a } _ {1}^{\dagger}\op{a}_{0}\bigr),    
\end{equation}
which amounts to disregard many-body interactions and keep (to first order) one-body terms, and from which many 
well-known system models can be recognized. For example, if $G_{01}=0$, $G_0=\omega_{0}$ 
and $G_{1}=\omega_{1}$, $\op{g}$ corresponds to the Hamiltonian of two independent 
harmonic oscillators of frequencies $\omega_{0}$ and $\omega_{1}$. If, instead, 
$G_{0}=G_{1}=0$ and $G_{01}\neq0$ is constant, $\hat g$ can be identified with the part of the 
Bose-Hubbard Hamiltonian related with the tunneling of particles between neighboring sites (with $|G_{01}|$ the intensity 
of the tunneling), or simply, the Hamiltonian that describes the transitions 
between the two levels considered.

Now, the Schwinger transformation \cite{Schwinger1952} maps the bosonic two-mode system into a system of $N$ elementary $1/2$-spins ($N$ qubits) with 
total 
angular momentum $\boldsymbol{\hat J}$. This allows us to perform our analysis resorting to angular momentum operators (and therefore applicable to 
angular momentum systems), defined in terms of the creation and annihilation operators as follows
\numparts
\beq\label{SchwingerTransform}
       \op{J}_{+}=\op{J}_{x}+i\op{J}_{y}=\op{a}_{0}^{\dagger}\op{a}_{1},\quad
       \op{J}_{-}=\op{J}_{x}-i\op{J}_{y}=\op{a}_{1}^{\dagger}\op{a}_{0},\quad
       \op{J}_{z}=\frac{1}{2}(\op{n}_{1}-\op{n}_{0}),
     \eeq
     where 
    \begin{equation}\label{Ji}
\hat J_l=\frac{1}{2}\big(\op{\sigma}_l\otimes \mathbb{I}_{2^{N-1}}+\mathbb{I}_{2}\otimes \op{\sigma}_l\otimes 
\mathbb{I}_{2^{N-2}}+\cdots+\mathbb{I}_{2^{N-1}}\otimes \op{\sigma}_l\big),
\end{equation} 
\endnumparts
$l=x,y,z$, $\mathbb{I}_{d}$ stands for the identity operator in a $d$-dimensional Hilbert space, and $\op{\sigma}_l$ denotes a Pauli operator, 
expressed in the basis $\{\ket{0},\ket{1}\}$ such that $\op{\sigma}_z\ket{0}=-\ket{0}$ and 
$\op{\sigma}_z\ket{1}=\ket{1}$. By use of the Schwinger's transform (\ref{SchwingerTransform}), \eref{1BodyG} can be written as
\begin{equation}\label{g}
    \op{g}=\frac{1}{2}(G_{0}+G_{1})\hat N+(G_{1}-G_{0})\op{J}_{z}+2G_{01}\op{J}_{x},
\end{equation}
with $\hat N$ the total number operator $\hat N=\op{n}_0+\op{n}_1$. 

The eigenvectors of $\hat g$, denoted as $\ket{g_n}$ with $n$ a quantum 
number that characterizes the state, correspond to a nondegenerate and equally-spaced 
spectrum, that is,
\begin{equation}\label{spectrum}
 \op{g}\ket{g_n}=g_n\ket{g_n},\quad {\rm with} \quad g_{n}=A+na,
\end{equation}
where $n=0,1,\ldots N$ and
\begin{equation}\label{Aa}
A=\frac{N}{2}\big(G_{0}+G_{1}-a\big),\quad 
a=\sqrt{(G_1-G_0)^{2}+4G^2_{01}}.
\end{equation}
For $G_{01}=0$, the common eigenvectors of $\op{J}_{z}$ and $\op{N}$, $\ket{N,n}$, correspond to the two-mode Fock states 
$\ket{n_0}\otimes\ket{n_1}=\ket{N-n}\otimes\ket{n}=\ket{N-n}\ket{n}$ that denote a state for which  $N-n$ particles dwell at 
level 0 and 
$n$ at level 1 (here $\ket{n_{i}}$ denotes an eigenvector of $\hat n_i$ such that $\hat n_i\ket{n_{i}}=n_i\ket{n_{i}}$). Thus, (for $G_{01}=0$) 
we get
\beq \label{gn}
\hat g\ket{g_n}=g_n\ket{g_n},\quad \ket{g_n}=\ket{N-n}\ket{n}, \quad g_n=A^0+na^0,
\eeq
with $A^0=G_0N$ and $a^0=G_1-G_0$, where we have assumed that $G_{1}>G_{0}$.
\subsection{Reaching orthogonality}
We have seen that the eigenvectors of the generator $\hat g$ given in (\ref{g}) constitute a set of nondegenerate states whose eigenvalues satisfy 
$g_n<g_{n+1}$. Therefore, any superposition of the form
\beq \label{psi0q3}
 \ket{\psi_0}=\sum^{N}_{n=0}\sqrt{r_n}e^{i\varphi_n}\ket{g_n}
\eeq
is amenable to the analysis presented in section \ref{speedlimit}, provided the set 
$\{r_n\}$ is such that the states that contribute to the sum are all equally probable and correspond to a set of equally-spaced eigenvalues. Thus, we 
will assume that there are 
$\mathcal{N}$ nonzero and equal coefficients $r_n$ (with 
$2\leq \mathcal{N}\leq N+1$), so that 
\beq \label{psi0q4}
 \ket{\psi_0}=\frac{1}{\sqrt{\mathcal{N}}}\sum_{k=1}^{\mathcal{N}}e^{i\varphi_{n_{k}}}\ket{g_{n_k}},
\eeq
where $\{n_k\}=\{n_{1},n_{2}\ldots,n_{\mathcal{N}}\}$ is a subset of equidistant integers of the set $\{n\}=\{0,1,\dots,N\}$, i.e., 
\beq
\Delta g=g_{n_{k+1}}-g_{n_k}=(n_{k+1}-n_{k})a=ma,
\eeq
with $m$ a fixed integer that satisfies $1\leq m\leq N$. This bound follows from the fact that $m$ is 
the difference between a pair of integers in the set $\{n\}$; however, it should be noticed that the specific value of $m$ is also restricted by the 
value of $\mathcal N$ (see equation (\ref{cotam}) below).

We can now resort to equation (\ref{gammapoli}) to conclude that the 
state (\ref{psi0q4}) evolving under 
the transformation $e^{-i\hat g\gamma}$ with $\hat g$ given by (\ref{g}), reaches an orthogonal 
state at
\beq \label{ortNqubit}
\widetilde\gamma=\frac{2\pi}{\mathcal{N}ma}.
\eeq
Now, the (equally probable) $\mathcal{N}$ states contributing to (\ref{psi0q4}) allow us to depict the state (\ref{psi0q4}) 
as an `$\mathcal{N}$-teeth comb' ---each tooth representing the state $\ket{g_{n_{k}}}$
with eigenvalue  $g_{n_k}$--- with teeth separation $ma$. The length of the comb is therefore
$L=ma(\mathcal{N}-1)$. The largest comb corresponds to the expansions (\ref{psi0q3}) that include the extremal states $\ket{g_0}$ and $\ket{g_N}$, and has the maximum length $L_{\max}=Na$. In its turn, the shortest combs correspond to the expansions that involve only two adjacent (minimally-spaced) states, $\ket{g_n}$ and $\ket{g_{n+1}}$, thus fixing the minimum length as $L_{\min}=a$ (see figure \ref{peine} below). From here it follows that 
\beq\label{cotam}
1\leq m(\mathcal N-1)\leq N.
\eeq
In terms of the length and the teeth spacing, equation (\ref{ortNqubit}) rewrites as 
\beq
\widetilde\gamma=\frac{2\pi}{L+ma}.
\eeq 
The maximal and minimal values of the length impose lower and upper bounds to $\widetilde\gamma$ in terms 
of $N$ and the single parameter of the comb $m$, as follows
\beq \label{ortNqubit2}
\widetilde\gamma_{l}\leq \widetilde\gamma\leq \widetilde\gamma_{s},
\eeq 
where
\beq \label{tminmax}
\widetilde\gamma_{l}=\frac{2\pi}{(N+m)a}\quad {\rm and} \quad 
\widetilde\gamma_{s}=\frac{\pi}{a}
\eeq    
correspond to the largest ($L_{\max}, m=1,2,\dots, N$) and the shortest ($L_{\min}, m=1$) comb, respectively. That is, larger 
combs reach orthogonality before shorter combs.
\begin{figure}[h]
\qquad\qquad\includegraphics[width=0.34\textwidth]{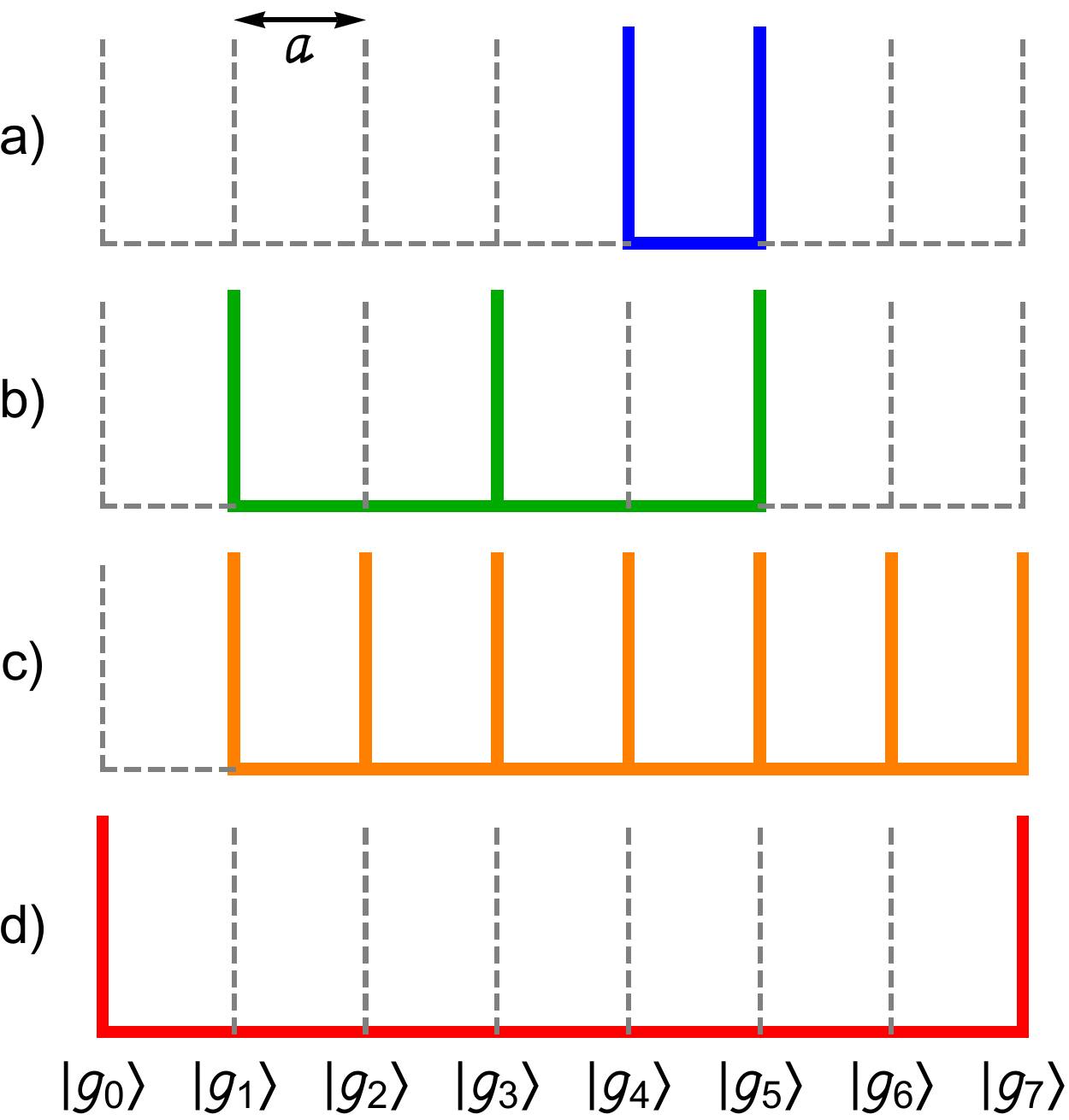}
 \caption{\label{peine} Comb representation of four possible states of the form (\ref{psi0q4}) for $N=7$. (a) The slowest state, for which $\widetilde\gamma=\pi/a$. The comb's parameters are $L=a$, $m=1$. (b) A state with 3 contributing $\hat g$-eigenstates, thus $\widetilde\gamma=\pi/3a$. The comb's parameters are $L=4a$, $m=2$. (c) A state with 7 contributing $\hat g$-eigenstates, and $\widetilde\gamma=2\pi/7a$. The comb's parameters are $L=6a$, $m=1$. (d) The fastest state, for which $\widetilde\gamma=\pi/7a$. The comb's parameters are $L=7a$, $m=7$. If the speed of evolution towards orthogonality is measured in terms of $\widetilde{\mathsf{g}}$, (c) would correspond to the slowest state, followed by (b), and then (a) and (d) would be the (equally) fastest states.}
\end{figure}

For fixed $N$, $\widetilde\gamma_{l}$ attains its 
lowest value for $m=m_{\max}=N$, whereas $\widetilde\gamma_{s}$ admits only $m=m_{\min}=1$. In both cases ($L=Na$ with $m=N$, and $L=a$ with $m=1$) we must have $\mathcal N=2$, precisely the value of $\mathcal N$ for which the minimum value $\widetilde\gamma_{\min}$ is reached. Therefore  
\beq \label{ineqs}
\widetilde\gamma_{\min}\biggl\vert_{m=N}=\frac{\pi}{Na}\leq\widetilde\gamma_{l}\leq \widetilde\gamma\leq 
\widetilde\gamma_{s}=\frac{\pi}{a}=\widetilde\gamma_{\min}\biggr\vert_{m=1}.
\eeq 
From equation (\ref{ineqs}) it follows that the `fastest' states 
are associated to the 2-teeth, largest comb with the largest possible separation, i.e.,
\beq\label{fastest}
\ket{\psi^{\rm fast}_0}=\frac{1}{\sqrt{2}}\left(\ket{g_{0}}+e^{i\phi}\ket{g_{N}}\right),
\eeq
with $\phi$ an arbitrary phase. Analogously, the `slowest' states are also represented by a 2-teeth comb, but now with the minimal teeth separation, i.e., by any of the states
\beq\label{slow}
\ket{\psi^{\rm slow}_0}=\frac{1}{\sqrt{2}}\left(\ket{g_{n_1}}+e^{i\varphi}\ket{g_{n_1+1}}\right),\quad g_{n_1+1}-g_{n_1}=a,
\eeq
with $n_{1}=0,1,\ldots N-1$. Figure \ref{peine} shows four possible comb arrays for a system with $N=7$. The gray (dotted) lines depict the 8 possible teeth the comb may have. From top to bottom, the figure represents four combs obtained by varying the number $\mathcal N$ of actual teeth and the separation $m$, whose corresponding values of $\widetilde\gamma$ are in decreasing order. In particular, the first comb saturates the last inequality in (\ref{ineqs}) ($\widetilde\gamma=\pi/a$) and therefore represents the slowest state $\ket{\psi^{\rm slow}_0}$, whereas the last comb saturates the left inequalities in (\ref{ineqs}) ($\widetilde\gamma=\pi/7a$) hence represents the fastest state $\ket{\psi^{\rm fast}_0}$. Notice that in terms of the relative, dimensionless parameter $\widetilde{\mathsf{g}}$, both states are equally fast, in line with the discussion below equation (\ref{ratio2}).

If the eigenvectors $\ket{g_n}$ are of the form (\ref{gn}) (which means no tunneling between sites, or no mode-interaction), then $\ket{\psi^{\rm fast}_0}$ is a balanced superposition of the Fock states 
$\ket{N}\ket{0}$ and $\ket{0}\ket{N}$, which, when written in the computational basis, is identified with the $N$-qubit Greenberger-Horne-Zeilinger 
(GHZ) state (with a relative phase)
\beq\label{ghz}
\ket{\psi^{\rm ghz}}=\frac{1}{\sqrt{2}}\big(\ket{0\cdots 
0}+e^{i\phi}\ket{1\cdots 1}\big).
\eeq
In its turn, in absence of tunneling, the state $\ket{g_{n_1}}$ in (\ref{slow}) is the Fock state $\ket{g_{n_1}}=\ket{N-n_1}\ket{n_1}$. Taking, for example, $n_1=0$, the state $\ket{\psi^{\rm slow}_0}$ written in the computational basis reads 
\beq \label{W}
\ket{\psi^{\rm w}}=\frac{1}{\sqrt{2}}\big(\ket{0\cdots 
0}+e^{i\varphi}\ket{W_N}\big),
\eeq
with $\ket{W_N}$ the $N$-qubit W state \cite{Dur2000}
\beq \label{WN}
\ket{W_N}=\frac{1}{\sqrt{N}}\big(\ket{100\cdots 
0}+\ket{010\cdots 
0}+\ket{00\cdots 
01}\big).
\eeq
The states $\ket{\psi^{\rm ghz}}$ and $\ket{\psi^{\rm w}}$ are thus extremal states that bound the value $\widetilde{\gamma}$ at which \emph{any} state of the form (\ref{psi0q4}), with $\ket{g_{n_k}}$ a two-mode Fock state, will evolve towards an orthogonal state. That is, \emph{none} of such states will attain an orthogonal state `faster' than $\ket{\psi^{\rm ghz}}$ nor `slower' than $\ket{\psi^{\rm w}}$. 
\subsection{Role of mode-entanglement in the non-tunneling regime}
Interestingly, the states $\ket{\psi^{\rm ghz}}$ (\ref{ghz}) and $\ket{\psi^{\rm w}}$ (\ref{W}) are representative of different types of \emph{multipartite entanglement} between $N$-distinguishable qubits \cite{Dur2000,Bengtsson2016,WalterMultipartiteEntanglement2019,GunePhysRep2009}. The connection between entanglement 
and the speed of quantum evolution goes back to the seminal work of Giovannetti et al. \cite{Giovannetti2003EPL,Giovannetti2003PRA}, that 
reveals that entanglement can `speed up' the evolution towards an orthogonal state when compared to separable (nonentangled) states. Since then, several
investigations exploring the effects of entanglement in the quantum speed limit problem have been advanced, for example, in
\cite{Batle2005,Borras2006,Borras2007,Kupferman2008,Frowis2012}.

In order to get further insight into the role of entanglement in the unitary evolution of composite systems towards orthogonality, instead 
of focusing on \emph{multipartite} entanglement between $N$ parties, here we pay 
attention to the \emph{bipartite} entanglement between the system's modes (or sites), as has been considered experimentally in 
\cite{KaufmanScience2016,CramerNComms2013}. Since 
our analysis involves pure states with only two modes ($0$ and 
$1$), the mode-entanglement can be directly computed resorting to the \emph{concurrence} \cite{Rungta2001}
\beq\label{concu}
C_{AB}=C(\ket{\psi}_{AB})=\sqrt{2\left(1-\Tr\rho^2_A\right)}=\sqrt{2\left(1-\Tr\rho^2_B\right)}\leq \sqrt{2(d-1)/d},
\eeq
which measures the (bipartite) entanglement between subsystems $A$ and $B$ when the whole composite $A+B$ is in the state $ \ket{\psi}=\ket{\psi}_{AB}
\in\mathcal H_A\otimes \mathcal H_B$. Here $d=\min\{\dim \mathcal H_A,\dim \mathcal H_B\}$, and $\rho_{A(B)}$ stands for the reduced density matrix of 
subsystem $A(B)$, i.e., $\rho_{A(B)}=\Tr_{B(A)}\bigl\{\ket{\psi}\bra{\psi}\bigr\}$. 

In the case where no interaction between modes occur ($G_{01}=0$), the eigenstates $\ket{g_{n_k}}$ in (\ref{psi0q4}) are the two-mode Fock 
states (\ref{gn}), and a straightforward calculation gives for the entanglement between the modes 
\beq \label{concmodes}
C_{01}=\sqrt{\frac{2(\mathcal N-1)}{\mathcal N}},
\eeq
which increases with $\mathcal N$ and attains its minimum value when $\mathcal N=2$, precisely when the Margolus-Levitin and Mandelstam-Tamm bounds are saturated. It thus 
follows that the extremal states $\ket{\psi^{\rm ghz}}$ and $\ket{\psi^{\rm w}}$ possess the \emph{same} and the \emph{minimum} value of entanglement 
between the modes. Comparison of equations (\ref{gammapoli2}) and (\ref{concmodes}) leads to an explicit relation between
$\widetilde{\gamma}$ and mode-entanglement, namely
\beq \label{concgamma}
\widetilde{\gamma}=\frac{\pi}{2(\langle g\rangle-g_1)}\,C^2_{01},
\eeq
meaning that for \emph{fixed} $\langle g\rangle-g_1$, the greater the amount of mode-entanglement the greater the `delay' in reaching an ortogonal 
state. Thus, if we consider a set of equally-weighted and spaced superpositions of Fock states, all having the same $g$-(relative)-expectation value, 
the first states that will attain an orthogonal state will be those with less entanglement, and correspond (in order of arrival at orthogonality) to combs with $2$, $3,\dots$, 
and so on, number of teeth. When the evolution is a Hamiltonian one, this means that for a given mean energy, or for the 
same energetic resources, mode-entanglement tends to slow down the evolution towards orthogonality.

If instead of fixing the (relative) mean value of $\langle g\rangle$ we fix the teeth separation $ma$, then 
$\widetilde{\gamma}$ and $C_{01}$ exhibit an inverse behaviour with respect to $\mathcal N$. The former is a decreasing function of $\mathcal N$ (see equation (\ref{ortNqubit})) whereas the latter increases with $\mathcal N$. Therefore, for the same spacing $ma$, the mode-entanglement favors a `faster' evolution to an orthogonal state.

\subsection{The effects of the transition between sites: $G_{01}\neq0$}
\begin{figure}[h]
\centering
 \includegraphics[width=0.75\textwidth]{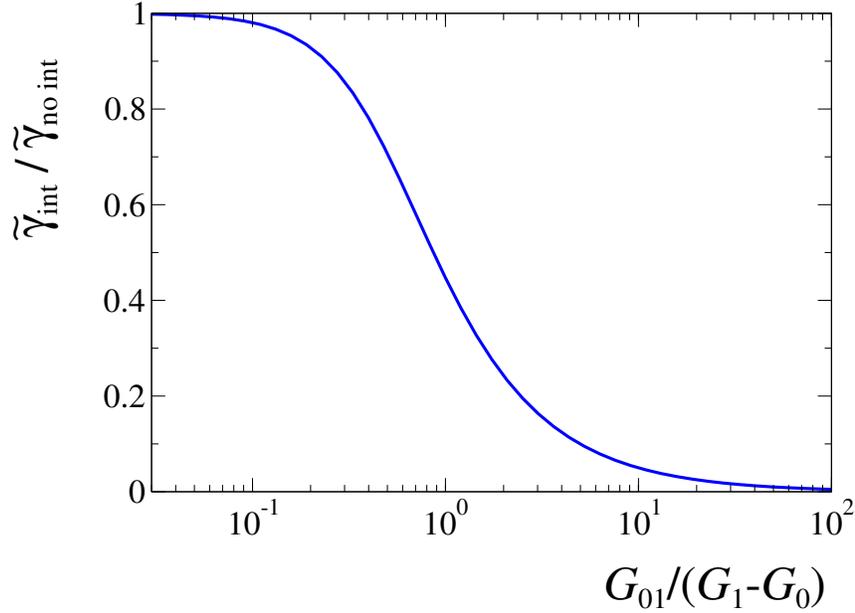}
 \caption{\label{Fig:gammaint}
$\widetilde{\gamma}_{\rm{int}}/\widetilde{\gamma}_{\rm{no\;int}}$ as a function of $G_{01}/(G_1-G_0)$. For fixed $G_0$, $G_1$, the value of 
$\widetilde{\gamma}$ decreases as the interaction gets stronger.
 }
\end{figure}

Resorting to equation (\ref{Aa}), we can compare the value of $\widetilde{\gamma}$ when the generator $\op g$ includes the mode-interaction terms 
($G_{01}\neq 0$), denoted as $\widetilde{\gamma}_{\rm{int}}$, with its value in absence of interaction ($G_{01}= 0$), represented by 
$\widetilde{\gamma}_{\rm{no\;int}}$. This gives 
\beq\label{gammaint}
\frac{\widetilde{\gamma}_{\rm{int}}}{\widetilde{\gamma}_{\rm{no\;int}}}=\Big[1+\frac{4G^2_{01}}{(G_1-G_0)^2}\Big]^{-1/2}\leq1,
\eeq
which shows that $\widetilde{\gamma}$ decreases as $G_{01}$ increases, or rather, 
that the transition between sites favors a \emph{faster} transit towards an orthogonal state, as depicted in figure \ref{Fig:gammaint} where the 
quotient in (\ref{gammaint}) is plotted as a function of $G_{01}/(G_1-G_0)$ (logarithmic scale).

The tunneling between sites will also affect the relation between $C_{01}$ and $\widetilde{\gamma}$ expressed in Eq. (\ref{concgamma}), where $G_{01}=0$ was assumed. Thus, in what follows we shall discuss the effects of finite values of $G_{01}$ in the 
connection between mode-entanglement and $\widetilde{\gamma}$.
We focus our analysis in the case when the initial state is
$\ket{\psi^{\rm{fast}}_{0}}$, given explicitly in (\ref{fastest}), for which $\widetilde{\gamma}=\pi/Na$ with $a$ given in (\ref{Aa}). Notice that
contrary to the $\gamma$-independent expression (\ref{concmodes}), $C_{01}$ will in general depend on $\gamma$ when $G_{01}\neq0$. 
In particular, it can be shown that the concurrence (\ref{concu}) for the $N$-boson state $\ket{\psi^{\rm{fast}}_{0}}$ is given by
\begin{equation}\label{Conc2}
      C_{01}^{\rm{fast}}=\left[2\left(1-\frac{1}{4}\sum_{n=0}^{N}\biggl\vert 
c_{n,0}+c_{n,N}e^{i(\phi-\pi\gamma/\widetilde{\gamma})}\biggr\vert^{4}\right)\right]^{1/2}\le\sqrt{\frac{2N}{N+1}},
     \end{equation}
where $c_{n,0}$, $c_{n,N}$ are the coefficients in the decomposition of the states $\ket{g_{0}}$ and $\ket{g_{N}}$, that define (\ref{fastest}), 
in the basis of the Fock states $\ket{N-n}\ket{n}$, i.e, $c_{n,0}=(\bra{N-n}\bra{n})\ket{g_0}$ and $c_{n,N}=(\bra{N-n}\bra{n})\ket{g_N}$. Such coefficients make $C_{01}^{\rm{fast}}$ to depend also on  
$G_{0}$, $G_{1}$ and $G_{01}$. For the sake of simplicity we set $G_{0}=0$ and use $G_{1}$ as a scale of units for $\op{g}$, so that  
$C_{01}^{\rm{fast}}$ depends only on $N$, $\gamma/\widetilde{\gamma}$ and $G_{01}/G_{1}$.

\begin{figure}
\centering
 \includegraphics[width=0.75\textwidth]{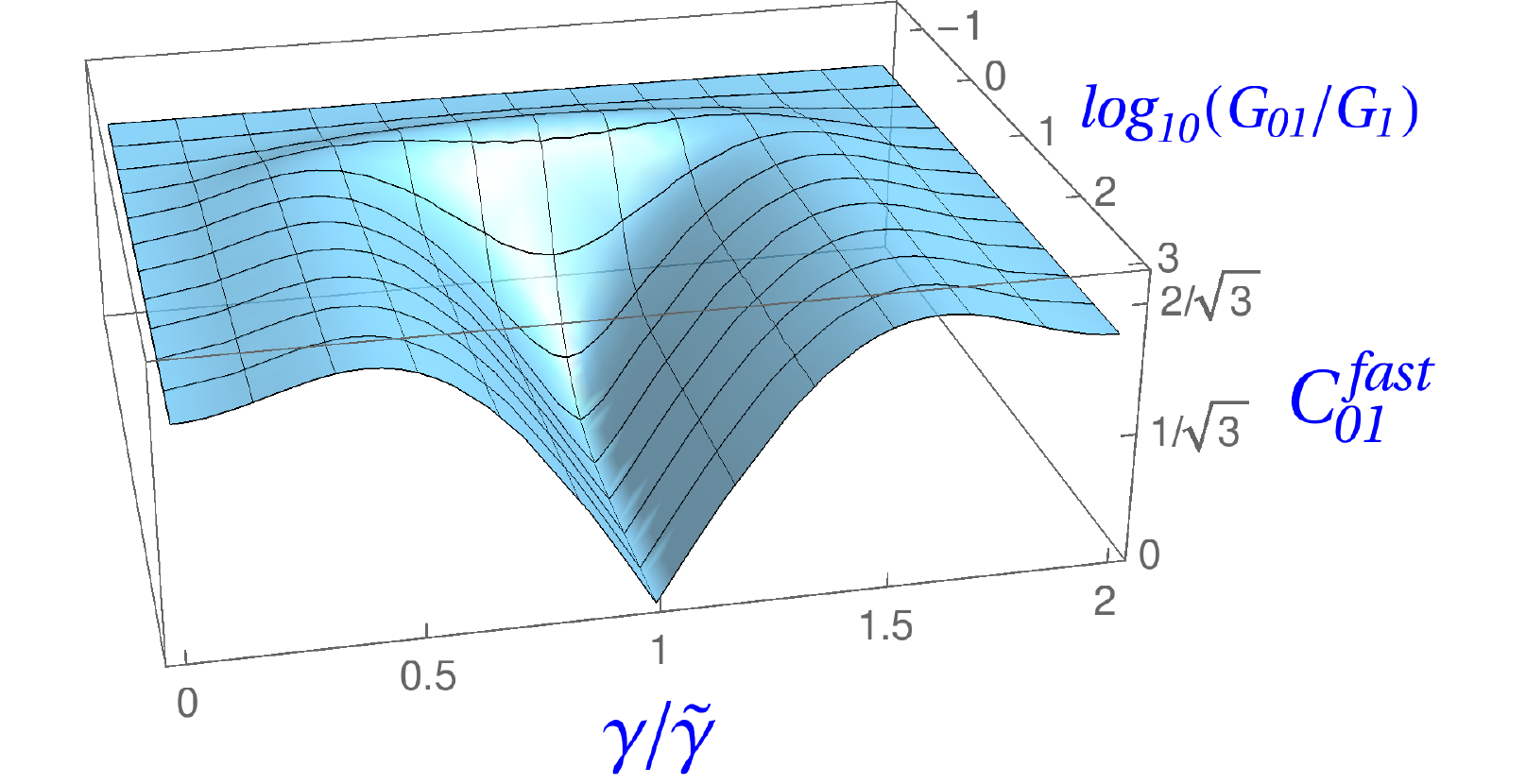}
 \caption{\label{Fig:Concurrence}Concurrence or mode-entanglement $C_{01}^{\rm{fast}}$ considering the initial state $\ket{\psi_{0}^{\rm{fast}}}$ for a system of two bosons as a function of the dimensionless 
parameter of the transformation, $\gamma/\widetilde{\gamma}$, and (the logarithm of) the dimensionless transition coefficient between modes, 
$G_{01}/G_{1}$ (for $G_{0}=0$).}
\end{figure}

In order to illustrate the effects of $G_{01}/G_{1}>0$ on the connection between $C_{01}^{\rm{fast}}$ and $\widetilde{\gamma}$, we put $N=2$ and 
$\phi=0$ for the sake of simplicity. The concurrence in this case has a maximum value equal to $2/\sqrt{3}\approx1.1547$, and is 
shown in figure \ref{Fig:Concurrence} as a function of $G_{01}/G_{1}$, and $\gamma/\widetilde{\gamma}$ (within the interval $[0,2]$, corresponding to a period of the dynamics as follows from equation 
\ref{PeriodGamma}). For $G_{01}=0$, $\ket{g_{0}}$ and $\ket{g_{2}}$ correspond, respectively, to the Fock states 
$\ket{2}\ket{0}$ and $\ket{0}\ket{2}$, and equation (\ref{Conc2}) reduces to $C_{01}^{\rm{fast}}=1$ for all $\gamma$, in agreement with equaion 
(\ref{concmodes}) (for $\mathcal N=2$), as expected. For small values of $G_{01}/G_{1}$, approximately for 
$G_{01}/G_1\lesssim0.3535$ (or $\log_{10} (G_{01}/G_1)\lesssim -0.4516$), $C_{01}$ is a concave function of $\gamma$ exhibiting a maximum of entanglement at $\gamma=\widetilde{\gamma}$, and the state (see equation 
(\ref{spectrum}))
\beq
\ket{\psi_{\widetilde{\gamma}}^{\rm{fast}}}=\frac{e^{-iA\widetilde{\gamma}}}{\sqrt{2}}\left(\ket{g_{0}}-\ket{g_{2}}\right)
\eeq
becomes maximally entangled, i.e.
$C(|\psi_{\widetilde{\gamma}}^{\rm{fast}}\rangle)=2/\sqrt{3}$, at $G_{01}/G_1\approx0.3535$. 

Interestingly, for $G_{01}/G_1\gtrsim0.3535$ the concurrence's maximum bifurcates into two local maxima at $\gamma_{\rm{m}}$ and  
$2\widetilde{\gamma}-\gamma_{\rm{m}}$, with $0<\gamma_{\rm{m}}<\widetilde{\gamma}$, while decreases monotonically with $G_{01}/G_1$ at 
$\gamma=\widetilde{\gamma}$. Thus, for large enough amplitudes of the transition between modes, 
$G_{01}$, two states emerge, namely $\ket{\psi_{\gamma_{\rm{m}}}^{\rm{fast}}}$ and $\ket{\psi_{2\widetilde{\gamma}-\gamma_{\rm{m}}}^{\rm{fast}}}$, that exhibit the corresponding maximum mode-entanglement (although slightly smaller than $2/\sqrt{3}$). In contrast, the mode-entanglement of the state 
$\ket{\psi_{\widetilde{\gamma}}^{\rm{fast}}}$ vanishes as $G_{01}/G_1\rightarrow\infty$. Indeed, a numerical analysis shows that in such a 
limit we have
\begin{equation}
 \ket{\psi_{\widetilde{\gamma}}^{\rm{fast}}}
\underset{G_{01}\to\infty}{
\longrightarrow}
-i\ket{1}\ket{1}=-i\ket{\Psi^{+}},
\end{equation}
with $\ket{1}\ket{1}$ a Fock state, whose mode-entanglement clearly vanishes. Notice that in the computational basis this state corresponds to the two-qubit Bell state $\ket{\Psi^{+}}=(\ket{01}+\ket{10})/\sqrt{2}$, with maximal qubit-qubit entanglement.

\section{Summary and final remarks}\label{fin}

We revisited the problem of determining conditions under which a pure state that evolves under an arbitrary unitary transformation reaches an 
orthogonal state in a finite amount of the transformation parameter. Simple geometric considerations disclose the existence of a `speed limit', and 
the necessity of dispersive dynamical variables ---specifically the variable associated to the generator of the evolution--- for the initial state to 
evolve towards a distinguishable one. In particular, when the evolution is generated by the Hamiltonian operator, these geometric observations go in 
line with the celebrated Mandelstam-Tamm bound. 

The geometrical picture presented here leads to consider the family of equally-weighted superpositions of, at least two, non-degenerate eigenvectors 
of the generator of the transformation, for which the evolution towards orthogonality is guaranteed. The distinguishable state is thus attained after 
`rotating' the initial state by a characteristic `angle' $\widetilde{\gamma}\Delta g$ (see equation (\ref{gammapoli})), and the generalized expressions 
(valid for arbitrary transformations) of the Mandelstam-Tamm and the Margolus-Levitin bounds are obtained. Moreover, for each state in the 
family under consideration, the relation $\sigma_{\op{g}}\le\mean{\op{g}}-g_{1}$ is satisfied, thus privileging the Mandelstam-Tamm bound over the 
Margolus-Levitin one. 

Some features of the dynamics are disclosed and the entropy production is analyzed, showing that in the simple model of a quantum clock advanced in \cite{AnandanPRL1990}, the relative entropy becomes maximal with each ticking of the clock, and for large $\mathcal N$ it does not change appreciable until the clock's hands point to the initial hour, after $(\mathcal N-1)$ tickings. In terms of the distinguishability of the states, this means that for a superposition of a sufficiently large number ($\gtrsim 10$) of (equally-weighted and spaced) non-degenerate states, the evolved state will remain almost distinguishable from the original one during the entire period $\Gamma$.  
  
Application of our generic analysis to multipartite systems allows, first, to contribute to the establishment and analysis of the relations between 
entanglement and the quantum speed limit, and second, to establish contact with physical systems that are customarily realized in the lab, such as 
bosons in optical lattices subject to a Bose-Hubbard Hamiltonian. In this respect, our results show that when no tunneling between sites exist, 
$\widetilde{\gamma}$ is directly proportional to the squared of the entanglement between the modes ($C_{01}$), so for $\langle \op g\rangle-g_1$ 
fixed, the higher the mode-entanglement is, the more amount of transformation is required to reach an orthogonal state. In contrast, for fixed $ma$, 
the higher the mode-entanglement is, the less amount of transformation is needed to attain a distinguishable state. Therefore, a high amount of 
mode-entanglement alone does not guarantee, nor prevents, a more `rapid' transit towards orthogonality. 

Now, when the bosons are allowed to tunnel between sites, a more complex relation between $\widetilde{\gamma}$ and $C_{01}$ is obtained, and the 
dynamics of the mode-entanglement becomes richer. This is exemplified for the initial state (\ref{fastest}), which corresponds to the superposition of 
eigenstates of the transformation generator with minimum and maximum eigenvalues. The observation of a more complex dynamics of $C_{01}$ is expected, since the 
transition between sites ---encoded in the `mode-interaction' part of the generator $\op g$--- typically affects the entanglement between modes.    

Our discussion on the relation between the mode-entanglement and the amount of transformation required to reach orthogonality, complements previous analysis 
\cite{Batle2005,Borras2006,Borras2007,Kupferman2008,Frowis2012} regarding the role of \emph{particle} entanglement on the quantum speed limit. As for the (multipartite) entanglement between the particles, here we 
found that in the nontunneling regime, the `slowest' ($|\psi^{\rm{w}}\rangle$) and the `fastest' ($|\psi^{\rm{ghz}}\rangle$) states correspond to 
paradigmatic states that exhibit multipartite entanglement \cite{Dur2000}. Further investigations will certainly reveal the more subtle relation between the quantum speed limit and the different types of multipartite entanglement.

There exist other geometrical approaches that bring out new and tighter bounds on the speed of evolution, such as that advanced in \cite{PiresPhysRevX2016}, and some comments on the main differences with the present treatment are in place. Here we make use of a rather simple mathematical framework which takes advantage of the fact that the overlap between an initial pure state and its evolved one is a complex scalar, a property that endows it with a two-dimensional representation and ultimately allows us to map the overlap into a set of (rotating) vectors in $\mathbb{R}^2$. Instead, other works analyze the distinguishability between two states (either pure or mixed) by focusing directly in their mutual `angle' (for instance the Bure or the Wigner-Yanase angle),  which naturally leads to a description involving differential geometry on Riemannian manifolds, where a family of metrics based on the geodesic distance between states is used to determine tighter quantum speed limits, that add to the Mandelstam-Tamm and Margolus-Levitin bounds. Considering if the approach presented here may be extended to recover those and other bounds is an interesting task that is left for future investigations, aimed at offering a simpler way of envisaging and constructing fundamental limitations in the evolution of dynamical systems.  

Within this latter aim, another line of research that also deserves further analysis is related to the universality of the limitations on the speed of evolution, as suggested by works showing that the speed limit remains in the transition from the quantum to the classical regime \cite{ShanahanPRL2018,OkuyamaPRL2018,ShiraishiPRL2018}. As stated before, the method followed in this paper is based on the mapping of the probability amplitude between two pure states into a vector in $\mathbb{R}^2$, whose direct classical counterpart remains unclear. There are, however, some directions within the same spirit of the present work that can be explored. For instance, the limit of continuous spectrum could be analyzed to provide some insights into the semi-classical limit; also an interesting approach would be to resort to the path functional description of the overlap. In a more ambitious scenario, some of the main ideas presented here and applicable to pure states only, may be extended to mixed states and then put into correspondence with probability distributions describing a classical system. All these open problems go beyond the scope of the present work, yet certainly are worthy of investigation.   


\ack
A.V.H. acknowledges financial support from DGAPA, UNAM through project PAPIIT IN113720 and 
F.J.S. acknowledges support by DGAPA-UNAM PAPIIT IN110120. The authors appreciate the comments of two  anonymous referees which helped to improve the manuscript.

\section*{References}
\providecommand{\newblock}{}

\end{document}